\documentclass[12pt]{article}

\usepackage[numbers]{natbib}
\usepackage{natbib}
\usepackage{amsmath}
\usepackage{amsfonts}
\usepackage{amssymb}
\usepackage{latexsym}
\usepackage{a4}
\usepackage{psfrag}
\usepackage{ulem}
\usepackage{epsfig}

\def\P{{\text P}}
\def\T{{\text T}}

\def \d{{\mathrm{d}}}
\def \R{{\mathbb{R}}}

\def \d{{\mathrm{d}}}

\def \LL{{\cal{L}}}
\def \HH{{\cal{H}}}
\def \PP{{\cal{P}}}

\def \II{{\cal{I}}}

\def \BB{\boldsymbol{B}}
\def \UU{\boldsymbol{U}}

\def \Bbeta{{\boldsymbol{\beta}}}
\def \bt{{\boldsymbol{\tau}}}

\begin{document}


\title{{\bf Generalized dynamics of moving dislocations in quasicrystals}}

\author{Eleni Agiasofitou$^{1,}$\footnote{{\it E-mail address:} agiasofitou@mechanik.tu-darmstadt.de}\, ,
Markus Lazar$^{1,2,}$\footnote{{\it E-mail address:} lazar@mechanik.tu-darmstadt.de}\, ,
and Helmut Kirchner$^{3,}$\footnote{{\it E-mail address:} kirchnerhok@hotmail.com}
\\  \\
$^{1}$ Emmy Noether Research Group, Department of Physics, \\Darmstadt University of Technology, \\Hochschulstr. 6, D-64289 Darmstadt, Germany\\
${}^{2}$
Department of Physics,
Michigan Technological University,\\
Houghton, MI 49931, USA\\
$^{3}$ INM - Leibniz Institute for New Materials, \\Campus D22, D-66123 Saarbr\"{u}cken, Germany
}

\date{\today}
\maketitle

\begin{abstract}
\noindent
A theoretical framework for dislocation dynamics in quasicrystals is provided according to the continuum theory of dislocations. Firstly, we present the fundamental theory for moving dislocations in quasicrystals giving the dislocation density tensors and introducing the dislocation current tensors for the phonon and phason fields, including the Bianchi identities. Next, we give the equations of motion for the incompatible elastodynamics as well as for the incompatible elasto-hydrodynamics of quasicrystals. We continue with the derivation of the balance law of pseudomomentum thereby obtaining the generalized forms of the Eshelby
 stress tensor, the pseudomomentum vector, the dynamical Peach-Koehler force density and the Cherepanov force density for quasicrystals. The form of the dynamical Peach-Koehler force for a straight dislocation is obtained as well. Moreover, we deduce the balance law of energy that gives rise to the generalized forms of the field intensity vector and the elastic power density of quasicrystals. The above balance laws are produced for both models. The differences between the two models and their consequences are revealed. The influences of the phason fields as well as of the dynamical terms are also discussed.
\end{abstract}


\section{Introduction}
Since the discovery of quasicrystals by \citet{Shechtman1984}, specimens of macroscopic size have been subjected to plastic deformation \citep{Bresson1994, Semadeni1997, Brunner1997}. Dislocation lines have been observed by means of the transmission electron microscope \citep{Wollgarten1993, Wollgarten1995}, the microstructure of deformed quasicrystals being similar to the microstructure of deformed crystals. The first direct evidence for dislocation motion in quasicrystals was observed by an {\it in-situ} straining experiment by means of electron microscope during the study of plastic deformation of icosahedral Al-Pd-Mn single quasicrystals by \citet{Wollgarten1995b}. Any theory of work hardening or work softening needs an understanding of the interaction between dislocations, and of dislocations with the applied stress. For crystals the essential ingredient is the force caused by a dislocation in the presence of stress and it is the well-known Peach-Koehler force \cite{Peach50}. As known in configurational mechanics \cite{Maugin93} the Peach-Koehler force density is the source term of the Eshelby stress tensor \cite{Eshelby75}. Obviously, a general expression for this configurational force caused by dislocations in quasicrystals is needed.
\par

In the literature, there exist different versions of
generalized linear elasticity theory of quasicrystals.
The difference of these versions lies in the dynamics of phonon and phason fields.
\citet{Bak85,Bak85b} argued that phason modes represent structural disorder or
structural fluctuations.
Following Bak's arguments in order to describe the dynamics of
phonons and phasons, \citet{Ding1993} and \citet{Hu2000}
implied that both phonons and phasons represent wave propagations.
For that reason, the equations of motion given by them are of wave-type for both, phonons and phasons, assuming furthermore that the phonons and phasons
have the same mass density.
Such a model is often called {\it elastodynamics of quasicrystals}.
Applications of elastodynamics of quasicrystals
are given by \citet{Fan1999b,Fan2003}, and \citet{Li2001} for the study of elastodynamics,
specific heat and thermodynamic functions.
\citet{Rochal2000,Rochal2002} generalized the elastodynamics of quasicrystals in that way that
the phason should possess a generalized phason density instead of the usual mass density of phonons. On the other hand, according to \citet{Lubensky85}, the dynamics of phasons represents a type of diffusion.
Lubensky's model \citep{Lubensky85} is usually called {\it hydrodynamics of quasicrystals}. However, according to this model, the dynamics of the phonon fields is also of diffusive character, possessing in this way at least the drawback
that classical elastodynamics cannot be recovered as a limit. A combination of the elastodynamics and the hydrodynamics of quasicrystals leads to a model which is called {\it minimal model of the phonon-phason elastodynamics} by \citet{Rochal2002}
and {\it elasto-hydrodynamic model of quasicrystals} by \citet{Fan2009}.
According to this model, that seems more appropriate for the description of the behavior of quasicrystals, the equations of motion for the phonons are equations of wave-type and for the phasons are equations of diffusion-type. In addition, several experimental results support this model as pointed out by \citet{Rochal2002}. Using coherent x-ray scattering, \citet{Francoual2003} have presented the first measurement of collective phason dynamics in quasicrystals, demonstrating that phason fluctuations are collective diffusive excitations. \citet{Zhu2008} used the elasto-hydrodynamic model for dynamic crack propagation in two-dimensional decagonal quasicrystals. However, as it is mentioned by \citet{Rochal2002}, there exists a class of incommensurate structures where
the collective phason modes are similar to sound ones and the elastodynamic model can be applied.
Thus, both the elastodynamic and the elasto-hydrodynamic model have physically realistic applications.
For that reason, we will deal
with {\it the elastodynamic} as well as {\it the elasto-hydrodynamic model
of quasicrystals} in the present paper.
\par
 A review, mainly from the experimental point of view, concerning electronic, transport, magnetic and mechanical properties of icosahedral and decagonal quasicrystals has been given by \citet{Takeuchi1994}. For the last twenty years, the elasticity theory of dislocations in quasicrystals has been an attractive research field \cite{Lubensky1986,Socolar1986, Ding1995b, Ricker2001, Zhu2007}.
For an overview on the (static)
elasticity theory of dislocations in quasicrystals we refer the reader to the articles of \citet{Hu2000}, \citet{Edagawa2001} and \citet{Edagawa2007}. The temperature is of fundamental importance in the mechanical behavior of quasicrystals as it was pointed out by \citet{Bresson1994}. \citet{Edagawa2007} report that generally quasicrystals are brittle at room temperature and dislocations within them are almost immobile. They can only be plastically
deformed at high temperatures above about $0.8\,{\text T}_m$ (${\text T}_m$: melting temperature). At sufficiently high temperature when atomic mobility by diffusion is faster than the dislocation velocity, perfect dislocations can migrate accompanying both phonon and phason strains.
 \par
The general expressions for the displacement fields (static case) induced by dislocations in quasicrystals have been calculated by \citet{Ding1995}. Also, \citet{Qin1997} have given the analytical expressions of the displacement fields (static case) induced by straight dislocations in decagonal, octagonal and dodecagonal quasicrystals.
As far as dynamics of dislocations in quasicrystals is concerned, using the elastodynamic model,
\citet{Fan1999} have found the analytical expressions for displacement and stress fields induced
by a moving screw dislocation in a one-dimensional hexagonal quasicrystal as well as the form of the energy.
Recently, using the elasto-hydrodynamic model of quasicrystals, \citet{Fan2009} have given analytical solutions
for a moving screw dislocation in some quasicrystalline systems. It should be noted that in the mentioned
works the plastic fields, e.g., plastic distortion, have been neglected in the equations of motion for dislocations in quasicrystals. However, plastic deformation is possible through dislocation motion at high temperatures, as observed
by \citet{Rosenfeld1995} and \citet{Bartsch2000}. Moreover, \citet{Li1999} have derived the Peach-Koehler force for a straight dislocation
in the static case for quasicrystals.
In the present paper, we derive the general expression of the Peach-Koehler force for
a moving dislocation in quasicrystals for both models.
\par
Based on compatible elasticity theory of quasicrystals, \citet{FM2004}
have defined the generalized Eshelby stress tensor for quasicrystals.
On the other hand, conservation laws in physical and material spaces for
a decagonal quasicrystal in compatible elastodynamics have been derived by Shi \cite{Shi2005}.
Shi \cite{Shi2007} has also derived the balance laws of pseudomomentum and scalar moment of momentum for an inhomogeneous material in the absence of body forces in the framework of the compatible elastodynamics of quasicrystals. Here, we find the balance laws of pseudomomentum and energy in incompatible elastodynamics as well as in incompatible elasto-hydrodynamics in presence of a continuous distribution of dislocations in quasicrystals.
\par
The paper is organized as follows: in Section 2 we set up the basic framework of moving dislocations in quasicrystals. Particularly, in the first subsection we give the geometric quantities of elastoplasticity of quasicrystals. We start with the compatibility conditions for the total distortion tensors and the total velocity vectors. The terms breaking the compatibility conditions lead us naturally to the definitions of the dislocation density and the dislocation current tensors in terms of the elastic and plastic fields, fulfilling the Bianchi identities. In the next subsection, we give the equations of motion for the elastodynamic as well as the elasto-hydrodynamic model of dislocations in quasicrystals. In the third section, we derive the balance laws that correspond to the infinitesimal variations of space and time, that is the balances of pseudomomentum and energy, respectively. The derivation of the dynamical Peach-Koehler force for quasicrystals possesses a prominent role. The above balance laws are obtained for both models. Eventually, the results are discussed in the last section and they are put into a broader context.

\section{The basic framework}
\subsection{Incompatible elasticity theory of quasicrystals}
An $(n-d)$-dimensional quasicrystal is defined as the projection of an $n$-dimensional periodic structure to $d$-dimensional space ($n>d$).
The  $n$-dimensional space can be decomposed into the direct sum of
two (orthogonal) subspaces, the $d$-dimensional {\it physical} or {\it parallel space} $E_\|$ and the $(n-d)$-dimensional {\it perpendicular space} $E_\bot$, that is, $\R^n=E_\|\oplus E_\bot$. In the
generalized elasticity theory of quasicrystals there are two types of fields, {\it the phonon} (conventional) {\it field}
and {\it the phason field}. Quantities which are associated with $E_\|$ are called {\it phonon-} and will be denoted by $(\cdot)^\|$ and quantities associated with $E_\bot$  are called {\it phason-} and will be denoted by $(\cdot)^\bot$. Apparently, the definitions of the physical quantities in quasicrystals are an extension of the definitions of the physical quantities in a usual medium enriched by the phason fields. It is important to say that all quantities depend only on
the physical space
coordinates ${\bf x} \in E_\|$. Thus, in addition to the usual {\it phonon displacement field} $u_i^\|({\bf x},t) \in E_\|$, there
exists {\it the phason displacement field}
$u_i^\bot ({\bf x}, t) \in E_\bot$. Phason displacement leads to local rearrangements of atoms, which are called {\it phason flips} and have a diffusive character \cite{Edagawa2007}.
\par
The displacement vector $\UU$ in the hyperspace $E_\| \oplus E_\bot$ is given in terms of the phonon and phason displacement fields as follows
\begin{equation}\label{U}
\UU=({\bf u}^\|, {\bf u}^\bot) \ \in E_\| \oplus E_\bot.
\end{equation}
{\it The total distortion tensors}, which are the gradients of the components of the displacement $\UU$ with respect to the spatial physical coordinates, and that correspond to the fields of phonon, ${\beta}^{\|\, \T}_{ij}$, and phason, ${\beta}^{\bot \T}_{ij}$,
can be decomposed into elastic $\beta^\|_{ij}, \ \beta^\bot_{ij} $ and plastic parts ${\beta}^{\|\, \P}_{ij}, \ {\beta}^{\bot \P}_{ij}$, respectively
\begin{alignat}{2}
&\label{up} u^\|_{i,j}:={\beta}^{\|\, \T}_{ij}=\beta^\|_{ij}+{\beta}^{\|\, \P}_{ij} \ \quad  &&\in E_\|\otimes E_\|,\\
&u^\bot_{i,j}:={\beta}^{\bot \T}_{ij}=\beta^\bot_{ij}+{\beta}^{\bot \P}_{ij} \quad &&\in E_\bot\otimes E_\|,
\end{alignat}
where comma denotes differentiation with respect to the spatial physical coordinates. Similarly, we can decompose the time derivative of the displacement fields
\begin{alignat}{2}
&\dot{u}_i^\|:=v_i^{\|\, \T}=v_i^\|+v_i^{\|\, \P}   \quad  &&\in E_\|,\\
&\label{vk}\dot{u}_i^\bot:=v_i^{\bot\, \T}=v_i^\bot+v_i^{\bot\, \P} \quad &&\in E_\bot,
\end{alignat}
where $v^\|_i$ and $v^\bot_i$ are {\it the elastic velocities} corresponding to the elastic fields of phonon and phason, respectively and
$v_i^{\|\, \P}$ and $v_i^{\bot\, \P}$ are {\it the plastic} or {\it initial} {\it velocities} \cite{Kossecka} in the spaces $ E_\|$ and $E_\bot$,
respectively. A superimposed dot denotes differentiation with respect to time.
\par
Due to the definition of the displacement field $\UU$ in the hyperspace (see eq.~(\ref{U})), the compatibility as well as the incompatibility conditions will
be analogous for the phonon and phason quantities. Therefore, the total fields satisfy {\it the compatibility conditions}
\begin{alignat}{2}
&e_{jkl}{\beta}^{\|\, \T}_{il,k}=0,\qquad
&&e_{jkl}{\beta}^{\bot \T}_{il,k}=0,\\
&\dot{\beta}^{\|\, \T}_{ij}-v_{i,j}^{\|\, \T}=0, \qquad&&\dot{\beta}^{\bot \T}_{ij}-v_{i,j}^{\bot\, \T}=0,
\end{alignat}
where $e_{jkl}$ is {\it the permutation tensor} or {\it Levi-Civita tensor}. On the other hand, the elastic as well as the plastic fields satisfy separately the following {\it incompatibility conditions}
\begin{alignat}{2}
&\alpha^\|_{ij}=e_{jkl}\beta^\|_{il,k}\qquad &&\in E_\|\otimes E_\|,\label{Comp Cond1}\\
&\alpha^\bot_{ij}=e_{jkl}\beta^\bot_{il,k}\qquad &&\in E_\bot\otimes E_\|,\label{Comp Cond2}\\
&I^\|_{ij}={\dot \beta}^\|_{ij}-v^\|_{i,j}\qquad &&\in E_\|\otimes E_\|,\label{Comp Cond3}\\
&I^\bot_{ij}={\dot \beta}^\bot_{ij}-v^\bot_{i,j}\qquad &&\in E_\bot\otimes E_\|,\label{Comp Cond4}
\end{alignat}
and
\begin{align}
&\alpha^\|_{ij}=-e_{jkl}{\beta}^{\|\, \P}_{il,k},\\
&\alpha^\bot_{ij}=-e_{jkl}{\beta}^{\bot \P}_{il,k},\\
&I^\|_{ij}=-\dot{\beta}^{\|\, \P}_{ij}+v_{i,j}^{\|\, \P},\\
&I^\bot_{ij}=-\dot{\beta}^{\bot \P}_{ij}+v_{i,j}^{\bot\, \P}.
\end{align}
The first two incompatibility conditions define {\it the phonon} and {\it the phason
dislocation densities}, $\alpha^\|_{ij}$ and $\alpha^\bot_{ij}$, respectively. The
last two incompatibility conditions define {\it the phonon} and {\it the phason
dislocation current tensors}, $I^\|_{ij}$ and $I^\bot_{ij}$ , respectively. The dislocation current tensor describes the
movement of dislocations and contains rate terms. One can read more about the significance and the
physical interpretation of the dislocation current tensor in \cite{Landau}. At this point it is necessary to generalize the well-known Bianchi identities \citep{Landau, Kosevich} for the dislocation density and the dislocation current tensors to quasicrystals. The dislocation density and the dislocation current tensors fulfill {\it the Bianchi identities}
\begin{alignat}{2}
&\alpha^\|_{ij,j}=0, \qquad   &&\alpha^\bot_{ij,j}=0,\label{Bianchi1}\\
&{\dot \alpha}^\|_{ij}=e_{jkl}I^\|_{il,k},\qquad
&&{\dot \alpha}^\bot_{ij}=e_{jkl}I^\bot_{il,k}.\label{Bianchi2}
\end{alignat}
\par
In a quasicrystal, a (perfect) dislocation is a line defect with the following
{\it Burgers vector} $\BB$ in the hyperspace $E_\| \oplus E_\bot $
\begin{equation}
\BB=({\bf b}^\|, {\bf b}^\bot) \ \in E_\| \oplus E_\bot,
\end{equation}
where
\begin{align}
b_i^\|=\oint_{\cal C}\beta^\|_{ij}\, \d x_j \quad \text{and} \quad
b_i^\bot=\oint_{\cal C}\beta^\bot_{ij}\, \d x_j
\end{align}
are the components of $B_i$ in the spaces $E_\|$ and $E_\bot$, respectively, and $\cal C$ is a closed contour surrounding the core of the
dislocation in the physical space $E_\|$. Alternatively, one can express the
Burgers vectors in terms of the dislocation density tensors using the Stokes theorem
\begin{align}
b_i^\|=\int_S \alpha^\|_{ij}\,\d S_j \quad \text{and} \quad b_i^\bot=\int_S \alpha^\bot_{ij}\,\d S_j,
\end{align}
where $S$ is the surface bounded by the curve $\cal C$ and pierced by the dislocation. If we integrate the Bianchi identities (\ref{Bianchi2}) over the surface $S$ and we use the Stokes theorem and the above formulas for the Burgers vectors, we obtain the following formulas
\begin{align}
\frac{\d b_i^\|}{\d t}=\oint_{\cal C}I^\|_{ij}\,\d x_j \quad \text{and} \quad \frac{\d b_i^\bot}{\d t}=\oint_{\cal C}I^\bot_{ij}\,\d x_j,
\end{align}
which are actually {\it conservation laws for the Burgers vectors}. The integrals on the right-hand side give the flux of the Burgers
vectors through the contour $\cal C$ per unit time and for that reason the dislocation current tensor is also called {\it dislocation flux
tensor} \cite{Landau}.
\subsection{Equations of motion of dislocations in quasicrystals}
We start our study with the elastodynamic model of quasicrystals.  {\it The Lagrangian density} $\LL$ is given in terms of {\it the kinetic energy density} $T$ and {\it the elastic energy density} $W$ as follows
\begin{equation}
\LL=T-W
\end{equation}
with
\begin{equation}\label{kin}
T=\frac{1}{2}\rho(v^\|_i)^2+\frac{1}{2}\rho_{\text {eff}}\,(v^\bot_i)^2,
\end{equation}
where $\rho$ is the usual mass density and $\rho_{\text {eff}}$ is the effective phason density \cite{Rochal2000, Rochal2002}. For the unlocked state\footnote{For the locked state there is a discussion whether the elastic energy density can have a quadratic form or the phasonic elastic energy density should depend linearly on the phason strains \cite{ Edagawa2001, Edagawa2007, Jeong1993, Trebin2006}.} of a quasicrystal, that means for $T>T_C$ ($T_C$: transition temperature) in 3D-systems \cite{Jeong1993}, the elastic energy density $W$ can be written in a quadratic form of phonon and phason strains
\begin{equation}\label{w}
W=\frac{1}{2}\beta^\|_{ij}C_{ijkl}\beta^\|_{kl}+\beta^\|_{ij}D_{ijkl}\beta^\bot_{kl}+\frac{1}{2}\beta^\bot_{ij}E_{ijkl}\beta^\bot_{kl},
\end{equation}
where $C_{ijkl}$ are the elastic constants in the classical elasticity theory, $E_{ijkl}$
are the elastic constants of the phason field in $E_\bot$ and $ D_{ijkl}$ are the elastic constants
associated with the phonon-phason coupling with the following symmetries
\begin{align}\label{Coef}
C_{ijkl}=C_{klij}=C_{ijlk}=C_{jikl},\quad D_{ijkl}=D_{jikl}, \quad E_{ijkl}=E_{klij}.
\end{align}
The specific form of the tensors of the elastic
constants and of the elastic energy density depends on the considered type of quasicrystals and has been examined for various types of quasicrystals for example pentagonal and icosahedral \cite{Levine1985},
octagonal and dodecagonal \cite{Socolar1989}, cubic \cite{Yang1993} as well as for two-dimensional quasicrystals \cite{Hu1996}.
\par
Here, the Lagrangian density $\LL$ is considered as a smooth function of ${\bf v}^\|, {\bf v}^\bot, \Bbeta^\|$ and $\Bbeta^\bot$. The volume
integral of $\LL$ is given by
\begin{equation}
\II=\int_\Omega \LL({\bf v}^\|, {\bf v}^\bot, \Bbeta^\|, \Bbeta^\bot) \d {\bf x},
\end{equation}
where $\Omega$ is an open, connected subset of $\R^3$ ($d=3$) with smooth boundary $\partial \Omega=S$. {\it The constitutive relations} for the considered Lagrangian density are
\begin{align}
&p_i^\|=\frac{\partial \LL}{\partial v^{\|}_i}=\rho v^\|_i, \label{CR1}\\
&p^\bot_i=\frac{\partial \LL}{\partial v^\bot_i}=\rho_{\text {eff}}\, v^\bot_i, \label{CR2}\\
&\sigma^\|_{ij}=\frac{\partial W}{\partial \beta^\|_{ij}}=C_{ijkl}\beta^\|_{kl}+D_{ijkl}\beta^\bot_{kl}, \label{CR3}\\
&\sigma^\bot_{ij}=\frac{\partial W}{\partial \beta^\bot_{ij}}=D_{klij}\beta^\|_{kl}+E_{ijkl}\beta^\bot_{kl} \label{CR4}.
\end{align}
Evidently, $p_i^\|$ and $p^\bot_i$ are {\it the phonon} and {\it phason momentum vectors}
and $\sigma^\|_{ij}$ and $\sigma^\bot_{ij}$ are {\it the phonon} and {\it phason stress tensors}, respectively. $\sigma^\bot_{ij}$ describes the stress components along the $x_i$
 direction in $E^\bot$ applied on the surface orthogonal to the $x_j$ direction in   $E^\|$. Notice that the phonon stress is symmetric,
$\sigma^\|_{ij}=\sigma^\|_{ji}$, but the phason stress is not, $\sigma^\bot_{ij}\neq\sigma^\bot_{ji}$, since the relevant constants $D_{ijkl}$ and $E_{ijkl}$ lack the necessary symmetries (see also, e.g., \cite{Ding1993}). Furthermore, the kinetic and the elastic energy densities, using the constitutive relations (\ref{CR1})--(\ref{CR4}), can be expressed as quadratic form
\begin{align}\label{TQ}
&T=\frac{1}{2}\bigl(p_i^\| v^{\|}_i+p_i^\bot v^{\bot}_i \bigl),\\
\label{WQ}
&W=\frac{1}{2}\bigl(\beta^\|_{ij}\sigma^\|_{ij}+\beta^\bot_{ij}\sigma^\bot_{ij}\bigl).
\end{align}
Introducing a {\it generalized (phason) body force density} $f_i^\bot$ and {\it the conventional
(phonon) body force density} $f_i^\|$, {\it the equations of motion}
 can be written in the form
\begin{align}\label{ME1}
&\dot{p}_i^\|-\sigma^\|_{ij,j}=f_i^\|,\\
\label{ME2}
&\dot{p}^\bot_i-\sigma^\bot_{ij,j}=f_i^\bot.
\end{align}
By substituting the constitutive relations (\ref{CR1})--(\ref{CR4}) in the above equations of motion  and making
use of the relations (\ref{up})--(\ref{vk}), we obtain {\it the inhomogeneous partial differential equations for the displacement fields} ${\bf u}^\|$ and ${\bf u}^\bot$ in a homogeneous material
\begin{align}
\label{u1-fe}
&\rho \ddot{u}_i^\|-C_{ijkl}u_{k,lj}^\|-D_{ijkl}u_{k,lj}^\bot=
\rho \dot{v}_i^{\|\, \P}-C_{ijkl}{\beta}^{\|\, \P}_{kl,j}-D_{ijkl}{\beta}^{\bot\, \P}_{kl,j}+f_i^\|,\\
\label{u2-fe}
&\rho_{\text {eff}}\, \ddot{u}_i^\bot-E_{ijkl}u_{k,lj}^\bot-D_{klij}u_{k,lj}^\|=
\rho_{\text {eff}}\, \dot{v}_i^{\bot\, \P}-D_{klij}{\beta}^{\|\, \P}_{kl,j}-E_{ijkl}{\beta}^{\bot\, \P}_{kl,j}+f_i^\bot,
\end{align}
where the plastic fields and the external forces are given as source terms. Eqs.~(\ref{u1-fe}) and (\ref{u2-fe}) are wave-type equations for the displacement fields ${\bf u}^\|$ and ${\bf u}^\bot$, respectively, for anisotropic media and they are coupled due to the constitutive tensor $D_{ijkl}$.
Eqs. (\ref{u1-fe}) and (\ref{u2-fe}) represent {\it the equations of motion for
the incompatible elastodynamics of quasicrystals}. In the case that $\rho_{\text {eff}}=\rho$ we recover the model that was proposed by \citet{Ding1993} and \citet{Hu2000}.
\par
We continue our study with the elasto-hydrodynamic model of quasicrystals. This model~\citep{Rochal2002} introduces the attenuation of phason modes which is essential
in the dynamical theory assuming that the phason bulk forces  $f_i^\bot=-D \dot{u}_i^\bot$,
where $D$ is the {\it friction coefficient}. In addition, the phason momentum
in quasicrystals is not conserved implying $\rho_{\text {eff}}=0$. Therefore, the phasons do not give a contribution to the kinetic energy. Under these circumstances, neglecting the phonon body force density $f_i^\|=0$
and putting for the incompatible case  $f_i^\bot=-D {v}_i^\bot$,
the equations of motions ~(\ref{ME1}) and (\ref{ME2}) take the form
\begin{align}\label{Minimal1}
\dot{p}_i^\|-&\sigma^\|_{ij,j}=0,\\
\label{Minimal2}
&\sigma^\bot_{ij,j}=D {v}_i^\bot.
\end{align}
The above equations can be written more analytically as follows
\begin{align}
\label{EM-minimal1}
&\rho \ddot{u}_i^\|-C_{ijkl}u_{k,lj}^\|-D_{ijkl}u_{k,lj}^\bot=
\rho \dot{v}_i^{\|\, \P}-C_{ijkl}{\beta}^{\|\, \P}_{kl,j}-D_{ijkl}{\beta}^{\bot\, \P}_{kl,j},\\
\label{EM-minimal2}
&D \dot{u}_i^\bot-E_{ijkl}u_{k,lj}^\bot-D_{klij}u_{k,lj}^\|=D {v}_i^{\bot\, \P}-D_{klij}{\beta}^{\|\, \P}_{kl,j}-E_{ijkl}{\beta}^{\bot\, \P}_{kl,j}.
\end{align}
The first eq.~(\ref{EM-minimal1}) is a wave-type one  for the displacement field ${\bf u}^\|$ while the second one (eq.~(\ref{EM-minimal2})) is a diffusion-type equation for the ${\bf u}^\bot$, both for anisotropic media.
Again the plastic fields are the source terms for the phonon and phason displacements and of course the two equations are coupled.
Eqs. (\ref{EM-minimal1}) and (\ref{EM-minimal2}) are {\it the equations of motion for
the incompatible elasto-hydrodynamics of quasicrystals}.
\section{Translational balance laws}
In this section, we derive the balance laws that correspond to the infinitesimal variations of space and time. It is clear
 that the translations in space will give the balance law of pseudomomentum while the variation of time will lead to the balance law of energy.
 We consider an arbitrary infinitesimal variation of the fields, leading to an infinitesimal variation of  $\II$
 \begin{align}\label{dI}
 \delta \II=\int_\Omega \delta \LL \, \d {\bf x}=\int_\Omega (\delta T-\delta W) \d {\bf x}.
 \end{align}

\subsection{Balance law of pseudomomentum--Peach-Koehler force}
We specify the variation here to the translational variation in the space coordinates as
\begin{equation}
\delta=\varepsilon_m \frac{\partial}{\partial x_m},
\end{equation}
where $\frac{\partial}{\partial x_m}$ is the generator of the translation group
in space and $\varepsilon_m$ are the parameters of the translation group. Then, eq.~(\ref{dI}) reduces to
\begin{equation}\label{dJ}
\delta \II=\int_\Omega \Bigl(\frac{\partial T}{\partial x_m}-\frac{\partial W}{\partial x_m}\Bigl)\varepsilon_m  \d {\bf x}.
\end{equation}
\par
We begin again our study with the elastodynamic model. Firstly, using the incompatibility conditions (\ref{Comp Cond3}) and (\ref{Comp Cond4}), the calculation of the gradient of the kinetic energy density  gives
\begin{equation}\label{pT}
\frac{\partial T}{\partial x_m}=\frac{\partial}{\partial t} \bigl(p_i^\| \beta^\|_{im}+p_i^\bot \beta^\bot_{im}\bigl)-{\dot p}_i^\| \beta^\|_{im}-
{\dot p}_i^\bot \beta^\bot_{im}-p_i^\| I^\|_{im}-p_i^\bot I^\bot_{im}.
\end{equation}
Also, the gradient of $W$, using the constitutive relations (\ref{CR3}) and (\ref{CR4}) and the symmetries of the
coefficients $C_{ijkl}$ and $E_{ijkl}$, gives
\begin{equation}\label{pW1}
\frac{\partial W}{\partial x_m}=\sigma^\|_{kl}\beta^\|_{kl,m}+\sigma^\bot_{kl}\beta^\bot_{kl,m}.
\end{equation}
Multiplying the compatibility conditions (\ref{Comp Cond1}) and (\ref{Comp Cond2}) with the permutation tensor, we obtain
two auxiliary identities which will be used in further calculations
\begin{align}\label{b}
\beta^\|_{kl,m}-\beta^\|_{km,l}=e_{mlj}\alpha^\|_{kj},\nonumber\\
\beta^\bot_{kl,m}-\beta^\bot_{km,l}=e_{mlj}\alpha^\bot_{kj}.
\end{align}
Next, if we substitute the relations (\ref{b}) in eq. (\ref {pW1}), we get
\begin{equation}\label{pW}
\frac{\partial W}{\partial x_m}=\sigma^\|_{kl}e_{mlj}\alpha^\|_{kj}+ \sigma^\|_{kl}\beta^\|_{km,l}
+\sigma^\bot_{kl}e_{mlj}\alpha^\bot_{kj}+\sigma^\bot_{kl}\beta^\bot_{km,l}.
\end{equation}
Inserting the eqs.~(\ref{pT}) and (\ref{pW}) into eq.~(\ref{dJ}) and rearranging the terms in order to
use the equations of motion (\ref{ME1})--(\ref{ME2}), we obtain the following
\begin{align}\label{dJfinal}
\delta \II=\int_\Omega \Bigl( &\frac {\partial} {\partial t}(p^\|_i\beta^\|_{im}+p^\bot_i\beta^\bot_{im})-\frac {\partial}{\partial x_l}
(\sigma^\|_{il}\beta^\|_{im}+\sigma^\bot_{il}\beta^\bot_{im})\\
&-p^\|_iI^\|_{im}-p^\bot_iI^\bot_{im}-e_{mlj}\sigma^\|_{il}\alpha^\|_{ij}-e_{mlj}\sigma^\bot_{il}\alpha^\bot_{ij}-
f^\|_i\beta^\|_{im}-f^\bot_i\beta^\bot_{im} \Bigl)\varepsilon_m  \d {\bf x}.\nonumber
\end{align}
On the other hand, we can also write
\begin{equation}\label{dL}
\delta \II=\int_\Omega \delta \LL\, \d {\bf x}=\int_\Omega  \frac{\partial \LL}{\partial x_m} \varepsilon_m \d {\bf x}=
\int_\Omega  \frac{\partial}{\partial x_l} (\LL\,\delta_{ml}) \varepsilon_m \d {\bf x}.
\end{equation}
Combining eqs.~(\ref{dJfinal}) and (\ref{dL}), we arrive at a {\it translational balance law for quasicrystals}
in global form
\begin{align}\label{BL}
&\int_\Omega \Bigl( \frac {\partial}{\partial x_l}
(-\LL\,\delta_{ml}-\sigma^\|_{il}\beta^\|_{im}-\sigma^\bot_{il}\beta^\bot_{im})+\frac {\partial} {\partial t}(p^\|_i\beta^\|_{im}+p^\bot_i\beta^\bot_{im})\Bigl)  \d {\bf x}\\
&=\int_\Omega \Bigl(p^\|_iI^\|_{im}+p^\bot_iI^\bot_{im}+e_{mlj}\sigma^\|_{il}\alpha^\|_{ij}+e_{mlj}\sigma^\bot_{il}\alpha^\bot_{ij}+
f^\|_i\beta^\|_{im}+f^\bot_i\beta^\bot_{im} \Bigl)  \d {\bf x}.\nonumber
\end{align}
In the above balance law we can recognize in the first integral two important quantities, the first one that enters in the divergence is {\it the Eshelby stress tensor}
\begin{equation}\label{Eshelby}
P_{ml}:=-\LL\,\delta_{ml}-\sigma^\|_{il}\beta^\|_{im}-\sigma^\bot_{il}\beta^\bot_{im}
\end{equation}
and the second one that enters in the time derivative is {\it the pseudomomentum vector}
\begin{equation}\label{pseudovector}
\PP_m:=-p^\|_i\beta^\|_{im}-p^\bot_i\beta^\bot_{im}.
\end{equation}
In elasticity, the tensor $P_{ml}$ is also called {\it the static energy-momentum tensor} \cite{Eshelby75} and is familiar in a general field theoretical concept \cite{Morse}. The integrand of the right-hand side integral is constituted by the sum of two forces,
{\it the dynamical Peach-Koehler force density}
\begin{equation}\label{Peach}
F^{\text {PK}}_m:=p^\|_iI^\|_{im}+p^\bot_iI^\bot_{im}+e_{mlj}\sigma^\|_{il}\alpha^\|_{ij}+e_{mlj}\sigma^\bot_{il}\alpha^\bot_{ij}
\end{equation}
and the {\it Cherepanov force density}
\begin{equation}\label{cherep}
F^{\text C}_m:=f^\|_i\beta^\|_{im}+f^\bot_i\beta^\bot_{im},
\end{equation}
which is produced by body forces in presence of elastic distortions. Both the Peach-Koehler force and the Cherepanov force are configurational forces and they are also called material forces \cite{Maugin93}. Their discussion, familiar from simple elastic media, can be taken over. For their elastic counterparts one can see the dynamical Peach-Koehler force in \cite{Schaefer, Lazar2008} and the Cherepanov force in \cite{Cherepanov1981, Kirchner1999}. After the definitions (\ref{Eshelby})--(\ref{cherep}), {\it the balance law of pseudomomentum} (\ref{BL}) can be elegantly written in global form as
\begin{equation}\label{EPglobal}
\int_\Omega \Bigl(\frac{\partial P_{ml}}{\partial x_l}-\frac{\partial\, \PP_m}{\partial t}\Bigl)\d {\bf x}=
\int_\Omega (F^{\text {PK}}_m+F^{\text C}_m)\d {\bf x}
\end{equation}
 or in local form
\begin{equation}\label{EPlocal}
\frac{\partial P_{ml}}{\partial x_l}-\frac{\partial\, \PP_m}{\partial t}=F^{\text {PK}}_m+F^{\text C}_m.
\end{equation}
It is evident from the obtained results, that
all physical quantities that enter into the balance law of
 pseudomomentum are strongly affected by the phason component of a dislocation. The balance law of pseudomomentum has been derived by Shi \cite{Shi2007} for inhomogeneous materials in absence of body forces in the framework of compatible elastodynamics of quasicrystals.
 \par
Furthermore, from the integral form of the balance law~(\ref{EPglobal}), we can obtain {\it the dynamical ${\boldsymbol J}$-integral}
\begin{align}\label{J}
J_m:=&\int_\Omega \Bigl(\frac{\partial P_{ml}}{\partial x_l}-\frac{\partial\, \PP_m}{\partial t}\Bigl)\d {\bf x}=
\int_S P_{ml}n_l \d S-\int_\Omega \frac{\partial\, \PP_m}{\partial t}\, \d {\bf x}\\
&=\int_\Omega (F^{\text {PK}}_m+F^{\text C}_m)\d {\bf x},\nonumber
\end{align}
where $\bf n$ is the  outward unit normal vector to the surface $S$.
\par
We proceed to find the form of the Peach-Koehler force for a straight dislocation. In analogy to the classical elastodynamics of dislocations in crystals \cite{Landau,Lardner}, we give the dislocation current and the dislocation density tensors for a straight dislocation to quasicrystals as follows
\begin{align}
\label{I}
&I_{ij}^\|=e_{jkl}V_k \alpha^\|_{il},\quad I_{ij}^\bot=e_{jkl}V_k \alpha^\bot_{il},\\
\label{a1}
&\alpha^\|_{ij}(x-x',y-y')=b_i^\|\tau_j \delta(x-x') \delta(y-y'),\\
\label{a2}
&\alpha^\bot_{ij}(x-x',y-y')=b_i^\bot\tau_j \delta(x-x') \delta(y-y'),
\end{align}
where ${\bf x}'=(x',y')$ is the position of the dislocation line and ${\bf V}={\dot{{\bf x}}}'$
is the velocity of the moving dislocation, $\bt=\bt({\bf x})\in E_\|$ is the tangent vector to the dislocation line and $\delta(\cdot)$ is the Dirac delta-function. Using the relations (\ref{I}) the Peach-Koehler force density (\ref{Peach}) becomes
\begin{equation}\label{Fak}
F^{\text {PK}}_m:=e_{mlj}\Bigl[(p^\|_iV_l+\sigma^\|_{il}) \alpha^\|_{ij}+(p^\bot_iV_l+\sigma^\bot_{il}) \alpha^\bot_{ij}\Bigl].
\end{equation}
Further, making use of the expressions (\ref{a1}) and \ref{a2}), eq. ({\ref{Fak}) gives the form of {\it the dynamical Peach-Koehler force for a straight dislocation in quasicrystals}
\begin{align}\label{PKsingle}
{\cal F}^{\text {PK}}_m&=\int\int_{\Omega'}e_{mlj}\Bigl[(p^\|_iV_l+\sigma^\|_{il})(x',y') \alpha^\|_{ij}(x-x',y-y')\nonumber\\
& \hspace{2.2cm}
+(p^\bot_iV_l+\sigma^\bot_{il})(x',y') \alpha^\bot_{ij}(x-x',y-y')\Bigl]\d x' \d y'\nonumber\\
&=\int\int_{\Omega'}e_{mlj}\Bigl[(p^\|_iV_l+\sigma^\|_{il})(x',y') b_i^\|+
(p^\bot_i V_l+\sigma^\bot_{il})(x',y') b_i^\bot \Bigl]\tau_j\, \delta (x-x') \delta (y-y')\d x' \d y'\nonumber \\
&=e_{mlj}\Bigl[(p^\|_iV_l+\sigma^\|_{il}) b_i^\|+
(p^\bot_i V_l+\sigma^\bot_{il}) b_i^\bot \Bigl]\tau_j.
\end{align}
It is obvious that the generalized dynamical Peach-Koehler force  of a moving dislocation with velocity ${\bf V}$ and Burgers vector $\BB$ in
quasicrystals is produced by the phonon and phason stresses and the phonon and
phason momenta.
If we omit the dynamical terms in eq.~(\ref{PKsingle}), then the obtained force is in accordance with the Peach-Koehler force that has been derived by \citet{Li1999}.
In addition, eq.~(\ref{PKsingle}) reduces to the classical dynamical Peach-Koehler force
for crystals (see, e.g., \citep{Kosevich}) when the phason fields are absent.
\par
In the case of the elasto-hydrodynamic model of dislocations, the general formulas of the balance law of pseudomomentum~(\ref{EPglobal}) or~(\ref{EPlocal}) are still valid. However, the quantities that constitute this balance law are modified as follows:
\begin{itemize}
\item[-]
{\it The Eshelby stress tensor} is given by the same formula~(\ref{Eshelby})
\begin{equation}
P_{ml}:=-\LL\,\delta_{ml}-\sigma^\|_{il}\beta^\|_{im}-\sigma^\bot_{il}\beta^\bot_{im}
\end{equation}
with a difference in the explicit form of the Lagrangian density $\LL=T-W$, because the kinetic energy density consists in the current case only of the phonon part, $T=\frac{1}{2}p_i^\| v^{\|}_i$.
\item[-]
{\it The pseudomomentum vector} is given by
\begin{equation}
\PP_m:=-p^\|_i\beta^\|_{im}.
\end{equation}
It is obvious that only the phonon fields give a contribution to the pseudomomentum vector, which has the same form as in incompatible elastodynamics of crystals.
\item[-]
{\it The dynamical Peach-Koehler force density} takes the following form
\begin{equation}
F^{\text {PK}}_m:=p^\|_iI^\|_{im}+e_{mlj}\sigma^\|_{il}\alpha^\|_{ij}+e_{mlj}\sigma^\bot_{il}\alpha^\bot_{ij}.
\end{equation}
\item[-]
{\it The Cherepanov force density} is given by
\begin{equation}
\label{Fc-phason}
F^{\text C}_m:=-D {v}_i^\bot \beta^\bot_{im}.
\end{equation}
The physical meaning of the Cherepanov force density is now a configurational
force density caused by the phason friction force ${\bf f}^\bot=-D\, {\bf v}^\bot$
in presence of the elastic phason distortion $\Bbeta^\bot$.
Thus, the term $D\,{\bf v}^\bot$ influences the ${\boldsymbol J}$-integral
only through the force density (\ref{Fc-phason}).
\end{itemize}
Furthermore, in the case of a straight dislocation the dynamical Peach-Koehler force density is given by the same formula as eq.~(\ref{PKsingle})
for $p^\bot_i=0$.
 \subsection{Balance law of energy}
 We continue in this fashion to find the balance law that corresponds to the time translation
\begin{equation}
\delta=\varepsilon \frac{\partial}{\partial t},
\end{equation}
where $\frac{\partial}{\partial t}$ is the generator of the translation group in
time and $\varepsilon$ is the parameter of the corresponding group. The corresponding variation of $\II$ will be
\begin{equation}\label{dIt}
\delta \II=\int_\Omega \Bigl(\frac{\partial T}{\partial t}-\frac{\partial W}{\partial t}\Bigl)\varepsilon\d {\bf x}.
\end{equation}
\par
We examine first the elastodynamic model. Due to the quadratic forms of $T$ and $W$ (eqs. (\ref{TQ}) and (\ref{WQ})), we have respectively
\begin{align}\label{Tt}
\frac{\partial T}{\partial t}=p_i^\| \dot{v}_i^\|+p_i^\bot \dot{v}_i^\bot
\end{align}
and
\begin{align}\label{wt1}
 \frac{\partial W}{\partial t}=\sigma^\|_{ij}\dot{ \beta}^\|_{ij}+\sigma^\bot_{ij}\dot{ \beta}^\bot_{ij}.
\end{align}
Using the compatibility conditions (\ref{Comp Cond3}) and (\ref{Comp Cond4}), eq.~(\ref{wt1})  can be further written
\begin{align}\label{wt2}
 \frac{\partial W}{\partial t}=\sigma^\|_{ij}I^\|_{ij}+\sigma^\|_{ij}v^\|_{i,j}+\sigma^\bot_{ij}I^\bot_{ij}+\sigma^\bot_{ij}v^\bot_{i,j}.
\end{align}
Inserting eqs. (\ref{Tt}) and (\ref{wt2}) in the expression (\ref{dIt}) and rearranging the terms in order to use also the equations of motion (\ref{ME1})--(\ref{ME2}), we finally obtain
\begin{align}\label{dIenergy}
\delta \II=\int_\Omega \Bigl( \frac {\partial} {\partial t}(p^\|_iv^\|_i+p^\bot_iv^\bot_i)-\frac {\partial}{\partial x_j}
(\sigma^\|_{ij}v^\|_i+\sigma^\bot_{ij}v^\bot_i)-\sigma^\|_{ij}I^\|_{ij}-\sigma^\bot_{ij}I^\bot_{ij}-f^\|_iv^\|_i-f^\bot_iv^\bot_i\Bigl)\varepsilon \d {\bf x}.
\end{align}
On the other hand,
\begin{equation}\label{dJL}
\delta \II=\int_\Omega \delta \LL\, \d {\bf x}=\int_\Omega \frac{\partial \LL}{\partial t}\varepsilon \d {\bf x}.
\end{equation}
Combining eqs.~(\ref{dIenergy}) and (\ref{dJL}), we obtain {\it the balance of energy for quasicrystals} in global form
\begin{align}\label{Benergy}
&\int_\Omega \Bigl( \frac {\partial} {\partial t}(p^\|_iv^\|_i+p^\bot_iv^\bot_i-\LL)-\frac {\partial}{\partial x_j}
(\sigma^\|_{ij}v^\|_i+\sigma^\bot_{ij}v^\bot_i)\Bigl)  \d {\bf x}\nonumber\\
&=\int_\Omega (\sigma^\|_{ij}I^\|_{ij}+\sigma^\bot_{ij}I^\bot_{ij}+f^\|_iv^\|_i+f^\bot_iv^\bot_i)  \d {\bf x}.
\end{align}
In the first integral, the quantity in the time derivative is {\it the Hamiltonian density}
\begin{equation}\label{Ham}
\HH=p^\|_iv^\|_i+p^\bot_iv^\bot_i-\LL=T+W
\end{equation}
and the quantity in the divergence is {\it the field intensity vector}
 \begin{equation}
S_j=\sigma^\|_{ij}v^\|_i+\sigma^\bot_{ij}v^\bot_i
\end{equation}
which describes the material energy flux. Finally, in the integral on the right-hand side we can see
{\it the elastic power density}
\begin{equation}\label{lamda}
\Lambda=\sigma^\|_{ij}I^\|_{ij}+\sigma^\bot_{ij}I^\bot_{ij}+f^\|_iv^\|_i+f^\bot_iv^\bot_i,
\end{equation}
which consists of two parts, the first one describes the energy exchange between the moving dislocation and the quasicrystal and the second one is the power produced by the body forces.
It can be seen that the elastic power density for a quasicrystal is produced
by phonon and phason stresses and phonon and phason body forces.
If we compare eq.~(\ref{lamda}) with the mechanical power in generalized continua (see, e.g., \cite{Jaunzemis}), we observe that the stress tensors have the same dimension as the dipolar forces and the dislocation current tensors play the role of the generalized velocities (velocity gradients).
\par
Taking into consideration the definitions (\ref{Ham})--(\ref{lamda}), {\it the balance law of energy} (\ref{Benergy}) can be rewritten  as
\begin{equation}\label{energy1}
\int_\Omega \Bigl(\frac{\partial \HH}{\partial t}-\frac{\partial S_{j}}{\partial x_j}\Bigl) \d {\bf x}=\int_\Omega \Lambda \, \d {\bf x}
\end{equation}
or in local form
\begin{equation}\label{energy2}
\frac{\partial \HH}{\partial t}-\frac{\partial S_{j}}{\partial x_j}=\Lambda.
\end{equation}
\par
In the same manner as in the previous subsection, we proceed to find the balance law of energy according to the elasto-hydrodynamic model of dislocations. The general formulas of the balance law of energy (\ref{energy1}) or~(\ref{energy2}) are still hold. However, the quantities that take part in this law are modified as follows:
\begin{itemize}
\item[-]
{\it The Hamiltonian density} changes due to the change of the kinetic energy density $T$
\begin{equation}
\HH=T+W, \qquad T=\frac{1}{2}\,p_i^\| v^{\|}_i.
\end{equation}
\item[-]
{\it The field intensity vector} does not change
 \begin{equation}
S_j=\sigma^\|_{ij}v^\|_i+\sigma^\bot_{ij}v^\bot_i.
\end{equation}
\item[-]
{\it The elastic power density} is modified as follows
\begin{equation}
\label{lambda-phason}
\Lambda=\sigma^\|_{ij}I^\|_{ij}+\sigma^\bot_{ij}I^\bot_{ij}-D (v^\bot_i)^2.
\end{equation}
The term $D\,{\bf v}^\bot$ enters the balance law (\ref{energy1}) through
the last term of the elastic power density (\ref{lambda-phason}).
\end{itemize}

\section{Conclusion}

We have embedded the dynamics of dislocations in quasicrystals in a general dynamic
field theoretical framework. For the quasicrystals the instruments of field theory
must be applied to an enlarged space, because the phason space is added to the
phonon space. Formally such space extension is akin to adding piezoelectricity,
piezomagnetics and magnetoelectricity to elasticity (see, e.g., \cite{Alshits}).
Procedurally the necessary extensions are straightforward.
\par
The dynamics of quasicrystals is a complex phenomenon and there is not yet an agreement, from the theoretical as well as from the experimental point of view, for the form of the generalized elasticity theory in dynamics of quasicrystals. In the present work, we cover two models that are mostly used in the literature for the description of the dynamic behavior of quasicrystals. The differences between these two models are essential since the governing equations for the elastodynamic model are of wave-type for both, phonons and phasons, while for the elasto-hydrodynamic model are of wave-type for the phonons and of diffusion-type for the phasons. The last years, the elasto-hydrodynamic model has gained the interest of the researchers since it seems that physically describes better the motion of quasicrystals. However, we should note that the equations of motion, corresponding to the last model, consist a coupled system of partial differential equations of wave-type and of diffusion-type thereby presenting some principal difficulties to be solved analytically.
\par
Furthermore, we have defined the dislocation density tensor and the dislocation current tensor
for quasicrystals in terms of the phonon and phason fields.
Using the field theoretical framework, we have introduced the Eshelby stress tensor,
the pseudomomentum vector, the Hamiltonian density and the field intensity vector
for a quasicrystal with incompatible fields.
By means of these quantities the dynamic translational balance laws are
established. As source terms of the translational balance laws we have obtained
configurational forces, namely the Peach-Koehler force and the Cherepanov force
in quasicrystals and the elastic power caused by phonon and phason stresses and
phonon and phason body forces.
\par
Finally, we would like to remind the reader that our results are valid at sufficiently
high temperatures since we deal with movement of dislocations.
For instance, one can see the exact range of temperature that can be used in
experiments as well as results from {\it in situ} straining experiments concerning
the dynamic behavior of dislocations for icosahedral Al-Pd-Mn single quasicrystals
in \cite{Messerschmidt2001}.

\section*{Acknowledgement} The authors wish to express their gratitude to the referees for their helpful remarks. The first two authors gratefully acknowledge the Emmy-Noether grant of the
Deutsche Forschungsgemeinschaft (Grant No. La1974/1-3).

\end{document}